\documentclass[12pt,preprint]{aastex61}
\renewcommand{\baselinestretch}{1.2}
\usepackage{graphicx}
\usepackage{amsmath,amssymb,latexsym}
\usepackage{multirow}
\usepackage{natbib}
\usepackage{url}
\usepackage{graphicx}

\usepackage{xcolor}
\usepackage{graphicx}
\usepackage{dcolumn}
\usepackage{bm}
\usepackage{comment}
\usepackage{hyperref}
\usepackage{tikz}
\usepackage{changepage}
\usepackage{braket}

\newcommand{\V}[1]{\mathbf{#1}}

\newcommand{\about}[1]{\sim\!\!{#1}}
\newcommand{\unit}[1]{\;\mathrm{#1}}
\newcommand{\grad}[1]{\nabla{#1}}
\newcommand{\di}{\partial}
\newcommand{\sub}[1]{_{\text{#1}}}

\newcommand{\mE}{M_{\oplus}}

\definecolor{nncolor}{rgb}{0.3,0.0,0.9}

\begin{document}

\title{The nature of gas giant planets}

\author{Ravit Helled}
\affil{Institute for Computational Science, University of Zurich\\
         Winterthurerstr.~190, CH- 8057 Zurich, Switzerland \\
         rhelled@physik.uzh.ch \\}
         \author{Naor Movshovitz}
          \affil{Dept.~of Astronomy and Astrophysics, University of California, Santa Cruz\\
           CA 95064, USA}
           \author{Nadine Nettelmann}
          \affil{Institute of Planetary Research, German Aerospace Center\\ 
          12489 Berlin, Germany}
         
\begin{abstract}
Revealing the true nature of the gas giant planets in our Solar System is
challenging.  The masses of Jupiter and Saturn are about 318 and 95 Earth
masses, respectively. While they mostly consist of hydrogen and helium, the
total mass and distribution of the heavier elements, which reveal information on
their origin, are still unknown. Recent accurate measurements of the
gravitational fields of Jupiter and Saturn together with knowledge of the
behavior of planetary materials at high pressures allow us to better constrain
their interiors.
Updated structure models of Jupiter and Saturn suggest that both planets have
complex interiors that include composition inhomogeneities, non-convective
regions, and fuzzy cores.
In addition, it is clear that there are significant differences between Jupiter
and Saturn and that each giant planet is unique. This has direct implications
for giant exoplanet characterization and for our understanding of gaseous
planets as a class of astronomical objects. In this review we summarize the
methods used to model giant planet interiors and recent developments in giant
planet structure models.
\end{abstract}

\section{Introduction}
The giant planets in the solar system are mysterious and complex objects. They
have been targets of detailed exploration from the ground and space for many
decades and their characterization remains a key goal of planetary science.
While significant progress in giant planet exploration has been made in the last
few years, in particular thanks to the Juno and Cassini spacecraft, new
questions have arisen, and there are many questions that still need to be
answered.

Constraining the composition of Jupiter and Saturn is of significant importance
due to several reasons. First, the composition of the planets can be used to
reveal information of the composition of the proto-planetary disk from which the
solar system formed. Second, exploration of Jupiter and Saturn allows for
comparative planetology which so we can understand whether Saturn is simply a
small version of Jupiter (spoiler: the answer is no) and this knowledge can be
reflected on the characterization of giant exoplanets. Third, the deep interiors
of giant planets are natural laboratories for materials at extreme pressure and
temperature conditions that cannot easily be achieved on Earth. Finally, a
determination of the composition and internal structure of the planets can be
used to constrain their formation and evolution histories.

In this chapter, we summarize the key methods used to model the structure of
giant planets and our current knowledge of the planets. Further information can
be found in several recent reviews on giant planet interiors, including
\cite{Fortney2010,Baraffe2014,Militzer2016,Guillot2015,Helled2018,Helled2019}
and references therein.

\section{Modeling Planetary Interiors\label{sec:modeling}}
For Earth, much of the information we have on its internal structure comes from
seismology. For giant planets, there is no way to directly inter their
compositions and internal structures, and therefore indirect methods must be
used.


\subsection{Mass and radius}
Before we discuss detailed structure models, it is worth noting that some
information on a planet's composition can be inferred from its mass and the
radius alone. Jupiter has a mass of $1.89818(9)\times{10}^{27}\unit{kg}$ and
Saturn's mass is $5.6834(2)\times{10}^{26}\unit{kg}$. Mass is relatively
easy\footnote{Rather, it's $GM$ that is easy to measure, often with exquisite
precision, and sometimes that is all we need. But here we really do need a mass,
in kilograms. A measurement of $G$, the universal constant of gravity, is
\emph{not} easy, and includes a small but non-negligible uncertainty.} to
measure (one or more natural satellites are helpful in that regard) with
reliable precision. Radius is trickier because the planets are not perfect
spheres. A simplification that will get us very close to the right volume is to
assume the planet's shape is an ellipsoid, estimate a surface radius at the
equator and the pole, and solve for the volume or, equivalently, a mean radius.
Jupiter has a mean radius of $69,911\pm{6}\unit{km}$, and Saturn
$58,232\pm{6}\unit{km}$.

\begin{figure}
\centerline{\includegraphics[width=0.7\textwidth]{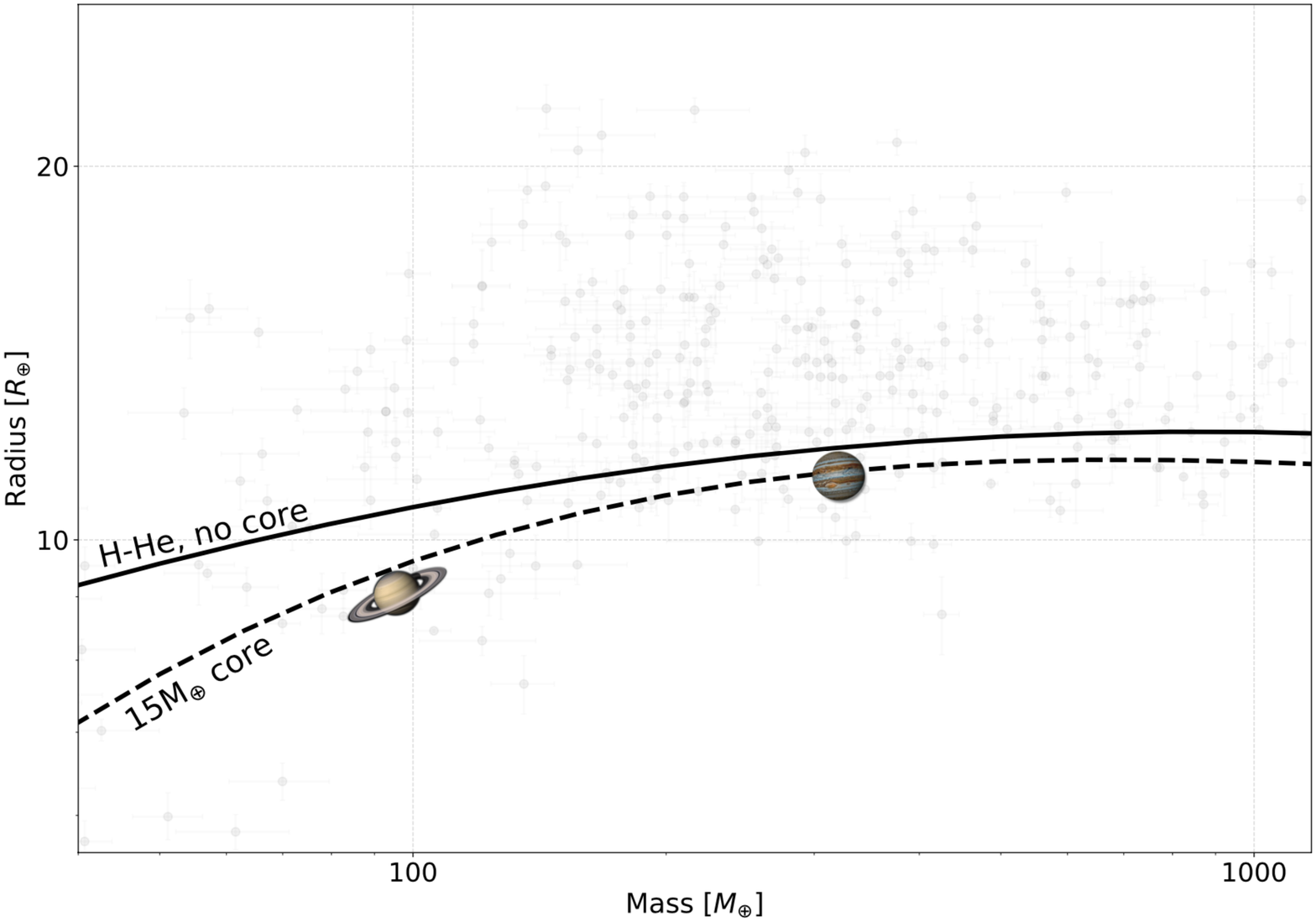}}
\caption{The mass-radius (M-R) relation of H-He-dominated planets. The solid
black curve corresponds to a H-He composition with a proto-soar ratio. The
dotted line demonstrates the influence of a $15\mE{}$ heavy-element core on the
M-R relation. Also shown are Jupiter and Saturn, and detected exoplanets in gray
(from the \emph{Dace} catalog (https://dace.unige.ch/dashboard/)).
\label{fig:M-R-HHe}}
\end{figure}

What do these values say about the planets' composition? We know that we are
dealing with, to first order, a mixture of hydrogen (H) and helium (He).  
But how much room is left
in the mix for heavier elements? Figure~\ref{fig:M-R-HHe} shows a theoretical
mass-radius relation for isolated H-He-rich planets. The solid curve assumes
pure H-He mixture with a protosolar ratio and the dashed curved is the same
mixture with a core of heavy elements. 
We see in the figure that both Jupiter and Saturn lie close to the H-He
curves, confirming that they consist of mostly hydrogen and helium, but well
below the curve for pure, protosolar H-He. This suggests that elements heavier
than H-He are present in significant quantity, although it is not
necessary that they are confined in a compact core. These heavier elements are
typically assumed to be ``rocks'' (i.e., silicates and sometimes metals) and/or
``ices'' (mainly H$_2$O but also CH$_4$ and NH$_3$).

In addition, since Saturn is farther from the pure-H-He curve than is Jupiter,
it is expected to be more enriched with heavy elements. This prediction might
appear to be contradicted by Saturn's lower average density
($687\unit{kg~m^{-3}}$ compared with Jupiter's $1326\unit{kg~m^{-3}}$). It's the
greater degree of compression of hydrogen and helium in Jupiter's interior, due
to its larger mass, that accounts for its higher overall bulk density.
\par 

\newpage
\subsection{Polytropic models}
A step beyond the mass-radius relation is the unit-index polytrope, the
relationship:
\begin{equation}
P=K\rho^2
\end{equation}
between pressure $P$ and density $\rho$. This simple and artificial assumption
can nevertheless be a surprisingly reasonable approximation of the
compressibility of a hydrogen-helium mixture at conditions typical of giant
planet envelopes, with the polytropic constant
$K=2.1\times{10}^{12}\unit{m^5\,kg^{-1}\,s^{-2}}$.

The polytropic relation and the condition of hydrostatic equilibrium
($dP/dr=-\rho(r)g(r)$, where $g(r)$ is the local acceleration due to gravity)
combine to an integro-differential equation on $\rho(r)$ which, with the
assumption of a spherical planet (i.e., neglecting oblateness due to rotation),
becomes the solvable ordinary differential equation
\begin{equation}
\frac{d}{dr}(r\frac{d\rho}{dr}) = -k^2r^2\rho(r),
\end{equation}
where $k=\frac{2\pi{G}}{K}$. Considering boundary conditions, the solution is
\begin{equation}
\rho(r)=\rho\sub{c}\frac{\sin(kr)}{kr}
\end{equation}
for $0\le{r}\le{R=\sqrt{\frac{\pi{K}}{2G}}}$.

The prediction then is that, neglecting the effects of rotation, a core, and
heavy elements, an isolated body in the giant planet mass regime should have a
predictable radius, which comes out to $R=70,300\unit{km}$. Apparently the
index-1 polytrope approximation is more appropriate for Jupiter than it is for
Saturn. This is both because the $P\propto\rho^2$ approximation doesn't fit
Saturn's present day envelope as well as Jupiter's, and also that Saturn's
interior is more enriched with heavy elements compared with Jupiter, as we
already suspected form the mass-radius relation.

More information on polytropic models, extended to account for rotation and
sometimes a core, can be found in \citep{Hubbard1975,Kramm2011,Paul2014}.

\section{Interior models}
While the mass-radius and polytropic relations can be used to infer some basic
predictions about the planetary bulk composition they provide no information on
the distribution of the materials. In order to constrain the density profile,
and therefore the material distribution, additional constraints are required.
For solar-system gas giants a critical measurement is their gravitational
fields. Additionally, the magnetic fields, 1-bar temperatures, atmospheric
composition, and static and dynamic rotation states are available and provide
further constraints. The key measurable properties of Jupiter and Saturn that
are used in interior models are listed in Table~\ref{tab:observables}.



Since giant planets consist of mostly fluid H and He, they do not
have solid surfaces below the cloud layers and as a result the ``surface'' of
the planet is defined as the location where the pressure is 1 bar, the pressure
at the Earth's surface. Often the measurement of the temperature at 1 bar is
used to infer the entropy of the outer envelope and therefore of the planetary
interior for adiabatic models where the temperature profile is set to be the
adiabatic gradient (see \cite{Militzer2016} and references therein for details).

\begin{table}[]
\center
{\begin{tabular}{@{}ccc@{}}
\toprule {\bf Property} & {\bf Jupiter} & {\bf Saturn}\\
\colrule
Distance to Sun (AU) & 5.204 & 9.582\\
Mass ($10^{24}\unit{kg}$) & $1898.13\pm{}0.19$ & $568.319\pm{}0.057$\\
Equatorial Radius (km) & 71492$\pm$4& 60268$\pm$4\\
Mean Density ($\mathrm{g/cm^3}$) & $1.3262\pm{}0.0004$ & $0.6871\pm{}0.0002$\\
Effective Temperature (K) & $124.4\pm{}0.3$ & $95.0\pm{}0.4$\\
1-bar Temperature (K) & $165\pm{}4$ & $135\pm{}5$\\
Rotation Period$^\text{i}$ & 9h 55m 29.56s & 10h 39m $\pm\sim$10m\\
$J_2 \times 10^6$ & $14696.572\pm{}0.014$ & $16290.557\pm{}0.028$\\
$J_4 \times 10^6$ & $-586.609\pm{}0.004$ & $-935.318\pm{}0.044$\\
$J_6 \times 10^6$ & $34.24\pm{}0.24$ & $86.334\pm{}0.112$\\
Love number $k_{2}$ & $0.565\pm{}0.018$ & $0.382\pm{}0.017$\\
\botrule
\end{tabular}
}
\caption{Basic observed properties of Jupiter and Saturn from \cite{Helled2018} and
references therein. Gravity field data from \cite{Iess2018,Iess2019}. The
gravitational coefficients correspond to the reference equatorial radii of
71,492 km and 60,330 km for Jupiter and Saturn, respectively. $^\text{i}$see  \cite{Helled2015,Fortney2018} for
discussion on Saturn's rotation rate uncertainty.}
\label{tab:observables}
\end{table}

An \emph{interior model} of a giant planet is a self-consistent solution of the
structure equations:
\begin{subequations}\label{eqs:struct}
\begin{align}
\frac{dm}{dr} &= 4\pi{}r^2\rho, \\
\label{eq:HE_omega}\frac{1}{\rho}\frac{dP}{dr} &= -\frac{Gm}{r^2}
+ \frac{2}{3}\omega^2r,\\
\label{eq:T}\frac{dT}{dr} &= \frac{T}{P}\frac{dP}{dr}\grad{T},
\end{align}
\end{subequations}
where $P$ is the pressure, $\rho$ is the density, $m$ is the mass inside a
pressure level of mean radius $r$, and $\omega$ is the rotation rate. The first
equation defines the transformation between a mass variable and radius variable.
The second equation is the condition of hydrostatic equilibrium, including a
correction term to account for oblateness due to uniform rotation. The third
equation describes the energy transport outward from the interior of the object
to its surface. The temperature gradient, $\grad{T}\equiv{}d\ln{T}/d\ln{P}$
depends on the energy transport mechanism. In convective regions the temperature
gradient is set to the adiabatic gradient
$\nabla\sub{ad}=\frac{\di\ln{T}}{\di\ln{P}}\arrowvert_S$, where $S$ is the
entropy. If the energy transport is by radiation, the \emph{diffusion
approximation} is
\begin{equation}
\nabla = \nabla\sub{rad} = \frac{3}{16\pi{}Gac}
\frac{\kappa_RL_rP}{mT^4} = \left(\frac{\di\ln{T}}{\di\ln{P}}\right)\sub{rad}.
\label{eq:radg}
\end{equation}
The derivative refers to the actual temperature--pressure variation in the
planetary structure of the planet, and $\kappa\sub{R}$ is the Rosseland mean
opacity. Often, radiation and conduction are treated together using an effective
opacity that accounts also for the contribution of conduction. Finally, a fourth
equation, the \emph{equation of state} (EoS), relates the density, pressure, and
temperature at each level. More details on the EoS are given in section 3.1.


The hydrostatic condition as expressed in eq.~(\ref{eq:HE_omega}) is only a
first order approximation of the effects of rotation. The precise statement of
the condition of hydrostatic equilibrium is this:
\begin{equation}
\frac{1}{\rho}\grad{P}=-\grad{U}
\end{equation}
where $U=V+Q$ is the sum of gravitational potential $V(\V{r})$ and centrifugal
potential $Q(\V{r})=-(1/2)\omega^2r^2\sin^2\theta$. 

The gravitational potential field $V$ is
\begin{equation}\label{Vgen_solu_Poiss}
V(\vec{r}) = -G\int d^3r' \:\frac{\rho(\vec{r'})}{|\vec{r'}-\vec{r}|},
\end{equation}
where the integration is over the, as yet unknown, volume of the planet. An
expansion in powers of $r$ reads
\begin{eqnarray}\label{eq:V3dim_Jn0}
	V(r,\,\phi,\,\theta) &=& -\frac{GM}{r}\left[
	\sum_{n=0}^{\infty} \left(\frac{r}{a_0}\right)^{\!\!-n}\!\!\! J_{n}\:P_n(\mu) \right.\nonumber\\
	+ && \left. \sum_{n=1}^{\infty}\sum_{m=1}^n \left(\frac{r}{a_0}\right)^{\!\!-n}
	\!\!\left(C_{nm}\:\cos m\phi \:+\: S_{nm}\:\sin m\phi\right) \:P_n^m(\mu) \right]
\end{eqnarray}
where $a_0$ denote the equatorial radius and $\mu=\cos{\theta}$. In this form
the information about the
planet's mass distribution is contained in the coefficients $J_n$, $C_{nm}$, and
$S_{nm}$. The \emph{zonal harmonics} $J_n$ are
\begin{equation}\label{eq:Jn}
Ma_0^nJ_n = -\int_{r'<R}d^3r'\rho{}r'^nP_n(\mu),
\end{equation}
with $P_n$ the Legendre polynomials. The \emph{tesseral harmonics} are
\begin{equation}\label{eq:Cnm}
Ma_0^nC_{nm} = 2\frac{(n-m)!}{(n+m)!}\int_{r'<R}d^3r'\rho(r')(r')^{n}
\cos{m\phi'}P_n^m(\mu'),
\end{equation}
and
\begin{equation}\label{eq:Snm}
Ma_0^nS_{nm} = 2\frac{(n-m)!}{(n+m)!}\int_{r'<R}d^3r'\rho(r')(r')^{n}
\sin{m\phi'}P_n^m(\mu'),
\end{equation}
with $P_n^m$ the associated Legendre functions. Often, the longitudinal
orientation of the planet-centered coordinate frame can be chosen so that
$S_{nm}=0$. If the gravity field does not dependent on longitude at all, i.e.,
if the field is symmetric with respect to the rotation axis, also $C_{nm}=0$. If
furthermore north-south symmetry with respect to the equatorial plane holds, the
odd $J_n$ vanish as well. Finally, for spherical planet, all $J_{n>0}=0$.

\begin{figure}
\centering
\includegraphics[width=0.650\textwidth]{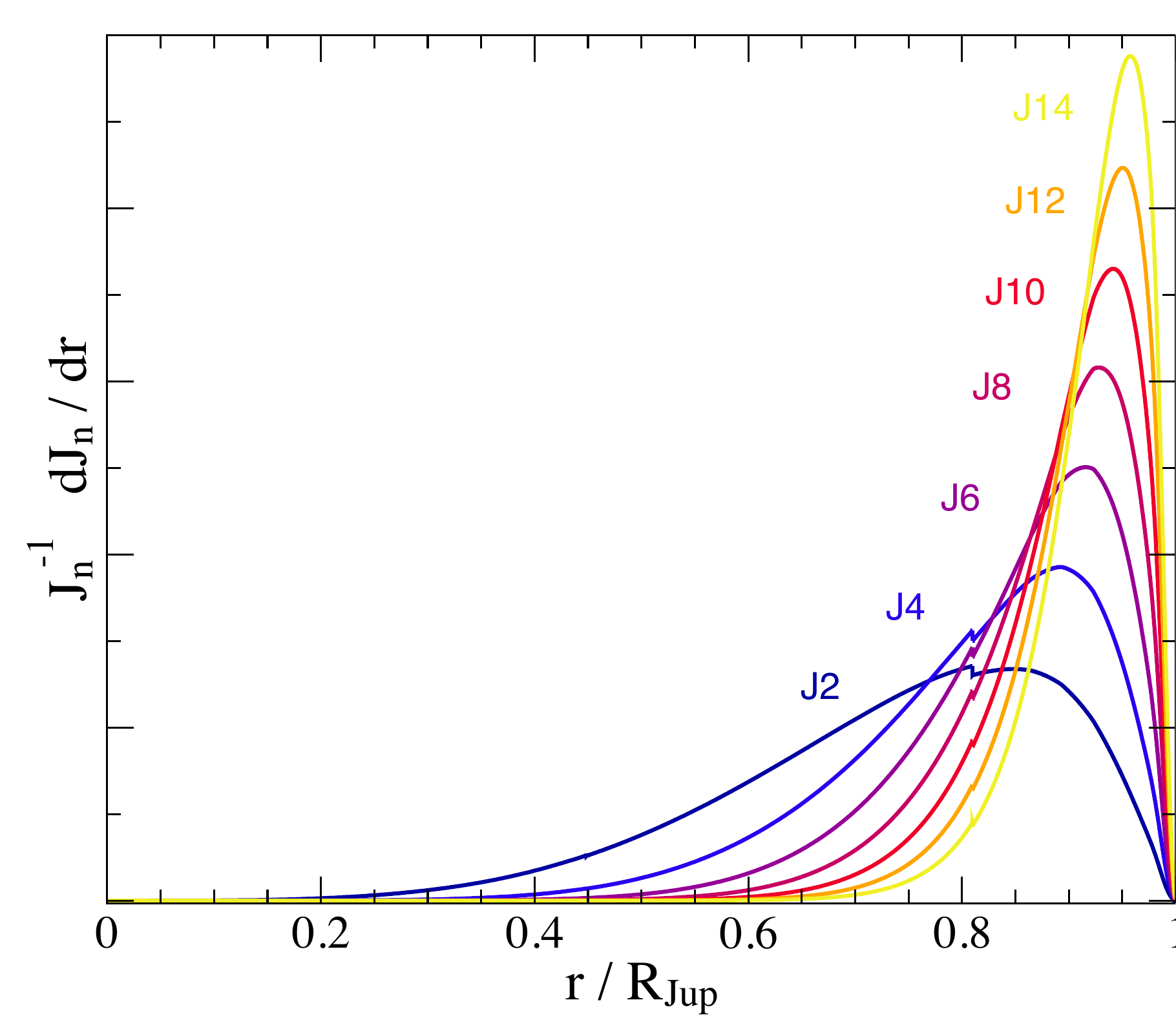}
\caption{\label{fig:nn_contribJ2n}
$J_{2n}$ contributions functions for a Jupiter model with Gaussian-Z in the deep
interior below 0.5 $R_{\rm J}$, where only $J_2$ is still sensitive. At the
small discontinuity seen in $J_2$ and $J_4$, an abrupt transition between
He-poor and He-rich regions occurs.}
\end{figure}

Due to rotation, giant planets are oblate, raising the even $J_{2n}$. Winds
perturb the density and thus introduce perturbation in the $J_n$. At first
glance, the zonal bands on Jupiter and Saturn appear north/south symmetric, but
a closer look at Jupiter shows this is not exactly the case. As a result, the
odd $J_n$ are raised on Jupiter \citep{Kaspi2018a}, and also Saturn
\citep{Iess2019}. One may thus write: 
\begin{equation}
J_n^\mathrm{obs} = J_n^\mathrm{static} + J_n^\mathrm{dyn}.
\end{equation}
The static gravitational harmonics $J_{2n}$ depend on the internal density
distribution, with the lower order harmonics $J_2,J_4$ being sensitive down to
$\sim 0.6\:R_{\rm J}$, while the static higher order harmonics probe the
density distribution in the deep atmosphere. This behavior is illustrated in
Fig.~\ref{fig:nn_contribJ2n} that shows an example of a Jupiter structure
model that fits the observed $J_2$ and $J_4$ values.

For both Jupiter and Saturn precise measurements of their $J_{n}$ values are
available from $J_2$ to $J_{12}$ \citep{Durante2020,Iess2019}. This allows to
constrain their internal structures far beyond what can be obtained by mass and
radius alone, as is the case for giant  exoplanets. 
On Jupiter and Saturn, the high-order $J_n$ are primarily influenced by the
winds, while the low-order moments $J_2$, $J_4$ primarily by the static field.
The latter are thus used to constrain interior models. Different interior models
differ in their prediction for $J_6$ and $J_8$
\citep{Guillot2018,Nettelmann2021} and therefore place somewhat different
constraints on the winds. Efforts to find an optimum model, which would yield
optimum agreement with the low-order field and the observed wind speeds are
still ongoing.

Eqs.~(\ref{eqs:struct}) cannot be inverted to yield a unique solution. Instead,
they are treated in a \emph{forward modeling} framework. A model is proposed, by
which we mean, some configuration of the planet's composition. 
Starting with a guess for $P(r)$, the temperature profile is calculated with
eq.~(\ref{eq:T}). The EoS is invoked to yield $\rho(r)$, then a
procedure\footnote{This is often a numerically sensitive and expensive
calculation \cite{Wisdom2016,Hubbard2014,Kong2013,Debras2017,Nettelmann2021}}
for calculating the potential, $V$. Finally the hydrostatic equilibrium equation
is integrated to update $P(r)$. The process iteratively approaches a
self-consistent solution, which is then checked against the known observables.

However, the solution is non-unique. The goal of such models could be a single,
best-fitting model, most consistent with every observed quantity, or an
exploration of all plausible models. Both approaches have been, and still are
actively pursued, as we discuss in the sections on Jupiter
(sec.~\ref{sec:jupiter}) and Saturn (sec.~\ref{sec:saturn}).

\subsection{The EoS of Hydrogen, Helium and Heavies}

As discussed above, in order to model the structure of giant planets one needs
to use an EoS in order to connect the density-pressure-temperature and other
thermodynamic properties.

Giant planet interiors serve as natural laboratories for studying different
elements at extreme conditions. In addition, calculating the EoS of materials in
Jupiter and Saturn interior conditions is a challenging task because molecules,
atoms, ions and electrons coexist and interact, and the pressure and temperature
range varies by several orders of magnitude, going up to several tens of
mega-bars (Mbar), or $\gtrsim{100}\unit{Gpa}$ and several $10^4$ Kelvins.
Therefore, information on the EoS at such conditions requires performing
high-pressure experiments and/or solving the many-body quantum mechanical
problem to produce theoretical EoS tables that cover such a large range of
pressures and temperatures. Despite the challenges, there have been significant
advances in high-pressure experiments and \emph{ab initio} EoS calculations.
More information on that topic can be found in
\cite{Fortney2010,Baraffe2014,Militzer2016,Guillot2015,Helled2018,Helled2020}
and references therein.

The composition of both Jupiter and Saturn is dominated by hydrogen and helium
(H, He), and therefore their modeling strongly depends on our understanding of
these materials at planetary conditions.

\subsubsection*{Hydrogen}
Our understanding of hydrogen at high pressures and temperatures is still
incomplete and is a topic of intensive theoretical and experimental research
(see \cite{Helled2020} and references therein). When it comes to giant planet
models, it has to be acknowledged that the EoS to be used must cover a large
range of temperatures (100 -- 10$^5$ K) and pressures (1 bar -- 10s Mbar, since
experiments are often limited to a specific range of parameters, planetary
scientists must rely on EoS calculations that are calibrated by experiments.


One of the most widely used EoS for H-He for giant planets and brown dwarfs is
the SCVH EoS \citep{Saumon1995}. This EoS based on an Helmholtz free-energy
model, from which thermodynamic parameters like pressure and entropy are derived
self-consistently. The SCvH EoS shows a first-order phase transition across a
wide range in temperature ($\sim10^3$--$2\times 10^4$K) and pressure
($\sim0.1$--100 Mbar), which was interpreted as a possible plasma phase
transition between molecular hydrogen and atomic, metallic hydrogen. This broad
phase transition region is smoothed by interpolation. Unfortunately, this region
corresponds to a region in the planetary deep interior that strongly affects the
inferred gravitational moments ($J_2$, $J_4$, see Fig.~\ref{fig:nn_contribJ2n}).

An alternative approach that is more challenging computationally, is to simulate
many-particle systems of electrons and ions, which basically obey the Coulomb
force, to infer the EoS and other material properties. However, just 1g of
hydrogen contains $5.98\times 10^{23}$ atoms, so approximations are still
required. Out of the several variants of approximations\cite{Helled2020}, the
method of DFT-MD simulations has been used to predict the EoS of H, He. In this
method, the ions are treated by classical Monte Carlo simulations (MD), and the
electrons by density functional theory (DFT). The electron wave functions and
eigenvalues are calculated using so-called pseudo-potentials, which approach the
Coulomb potential at sufficiently large interparticle distances to avoid strong
oscillations at small distances. This method is particularly suited for high
densities where Free-energy-based EoS show their largest uncertainties.

At present, there are three DFT-MD data based H EoS that are commonly used in
the planetary community: MH13 EoS \citep{Militzer2013}, REoS \cite{Becker2014},
and CMS19-EoS \cite{Chabrier2019}. The MH13 EoS uses DFT-MD calculations and is
calculated for a protosolar H-He mixture. It also provides the entropy. At lower
densities, this EoS is connected to the SCvH EoS ensuring a smooth transition in
entropy (and pressure). The R-EoS for H includes DFT-MD data between 0.1 and 10
g/cc.

\begin{figure}[tb!]
\centerline{\includegraphics[width=0.7\textwidth]{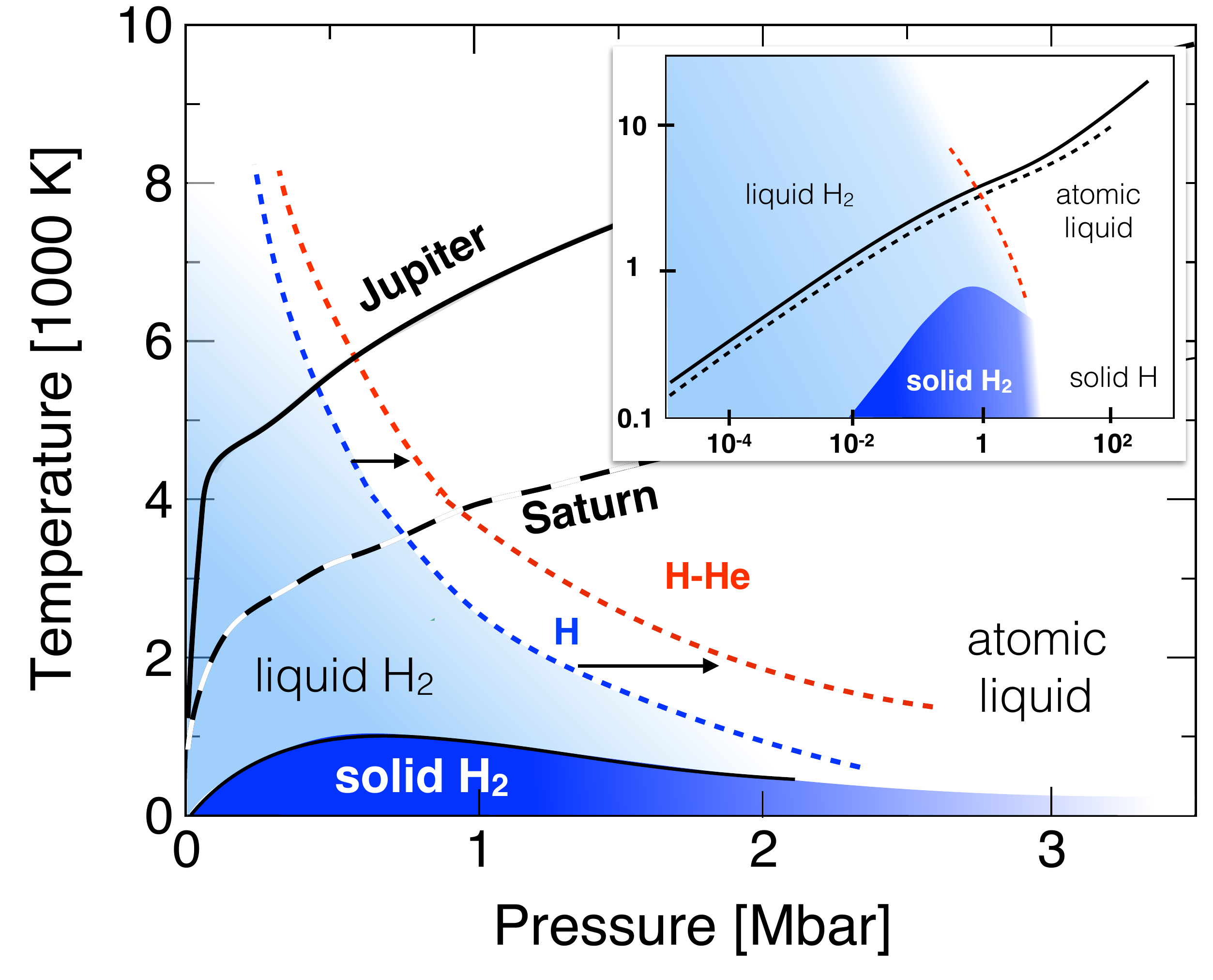}}
\caption{The phase diagram of hydrogen. {\bf Large panel:} A zoom to the
temperature-pressure region associated with hydrogen metallization. The blue and
red dashed curves present the theoretical prediction of the fluid metallization
of pure H  and of a H-he mixture as inferred from QMC simulations
\cite{Mazzola2018}. Also shown are examples of pressure-temperature profiles for
Jupiter (solid black curve) and Saturn (dashed curve). {\bf Small right panel:}
The H phase diagram on a larger pressure-temperature scale, showing the
different phases of H, and representative solutions for  Jupiter and Saturn. The
image is modified from \cite{Mazzola2018,Helled2020}.}
\label{fig:HHe_phase}
\end{figure}

Entropy values or adiabats are calculated using standard thermodynamic relations
\cite{Scheibe2019}. In the CMS19-EoS, DFT-MD data includes a small range of
0.3--5 g/cc and at lower densities, the SCvH EoS is employed. As a result of the
DFT-MD data in these three EoSs, the molecular-metallic transition occurs at
lower temperatures of a few 1000 K (see Fig.~\ref{fig:HHe_phase}). For a
comparison with experimental data and detailed discussions on EoSs we refer the
reader to the cited literature.
It was recently found that  interior models  of Jupiter and Saturn based on the DFT-MD H-He EoS data have a rather compressible adibat, meaning higher densities than for SCvH EOS at pressures where $J_4$ is sensitive.  As a result, DFT-MD based models yield low metallcities in the outer envelopes of Jupiter and Saturn.

\subsubsection*{Hydrogen-Helium}
The second most common element in giant planets is Helium. Experimental data and
numerical calculations of the phase diagram of H- He mixtures are now being
performed. For giant planet interiors, a key feature of such a mixture is the
immiscibility of He in H which is expected to lead to  He settling, or ``helium
rain'' \cite{Stevenson1977,Stevenson1977a,Morales2013,Brygoo2021}. The process of helium rain affects the cooling as well as the structure
of giant planets \cite{Pustow2016,Mankovich2019a}. The
conditions for de-mixing are still debated but it seems that Saturn's interior,
due to its lower temperature, is more affected by this process in comparison to
Jupiter \cite{Schottler2019}. Understanding the behavior of a H-He mixture is
also important since the existence of helium in H stabilizes the H molecules,
and was found to delay the H metallization towards higher densities
\cite{Mazzola2018}.

\subsubsection*{Heavy elements}
In astrophysics, heavy elements also known as ``heavies" correspond to all
elements that are heavier than hydrogen and helium. The heavy-element mass fraction is  often represented by the letter ``Z". Heavy elements in
giant planets can be metals (e.g., iron), silicates (e.g., silicon) or volatiles
(e.g., oxygen, ammonia). 
Although understanding the internal structure of
Jupiter and Saturn strongly depends on the H, and H-He EoS, a determination of
the total mass of heavy elements and how they are distributed within the
planetary interior is critical for our understanding of the formation and
evolution of the giant planets (see \cite{Helled2018} and references therein for
details). 

\subsubsection*{Empirical EoS}
An alternative approach to the EoS-based models described above is to
parameterize the $\rho(r)$ configuration rather than the H-He-Z mixture
configuration. The hydrostatic equation~(\ref{eq:HE_omega}), the temperature
equation~(\ref{eq:T}), and the equation of state are dropped from the iterative
solution cycle (although they can still be integrated after the fact) and the
solution is required only to match the known mass, radius, and gravity field
(and boundary conditions). This framework (which have long been termed
``empirical'') seems like a step backwards from EoS-based models but the goal is
different. While EoS-based solutions create very detailed, realistic, and easy
to interpret model planets, they invariably rely on a host of assumptions, some
made with sound physical reasoning and some made out of necessity, and thus are
limited in the types of solutions that can be produced. Of course,
composition-agnostic models are subject to their own assumptions and
simplifications, required to parameterize the density profiles, but they are not
subject to the same assumptions as EoS-based models, and in particular they are
not required to assume homogeneous composition layers anywhere in the planet.

In this way, composition-agnostic models can be used to explore a wider range of
plausible interiors. Examples of this approach applied to Saturn
\cite{Movshovitz2020} and Jupiter \cite{Neuenschwander2021} are discussed in the
respective sections. This approach also has been, and is being, applied to the
Uranus and Neptune \cite{Marley1995a,Podolak2000,Helled2011}.


\newpage
\section{Jupiter\label{sec:jupiter}}
Our understanding of Jupiter's interior has been challenged when Jupiter's
gravitational field was accurately measured by the {\it Juno} spacecraft
\cite{Iess2018}. Jupiter structure models pre-Juno era can be found in \cite{Miguel2016,Militzer2016}.  

The much more accurate gravity data has led to the development of more complex
structure models that include composition gradients and account for the
dynamical contributions. Interestingly, the improvements in the quality of the
gravity data led to many new questions regarding the planetary interiors, and
standard structure models have been challenged. 

Updated structure models that fit Juno data imply that Jupiter's interior is
characterized by a non homogeneous distribution of the heavy elements, where the
deeper interior is more metal-rich \cite{Wahl2017a,Debras2019}. In addition, in
these updated models Jupiter's core is no longer viewed as a pure heavy-element
central region with a density discontinuity at the core-envelope-boundary, but
as a central region enriched with heavies but also consists of lighter elements
such as H-He.
Such a fuzzy/diluted core could extend to a few tens of percents of Jupiter's
total radius. The new Jupiter structure models also imply that the temperature
profile in the planetary deep interior can significantly differ from the
adiabatic one \cite{Vazan2018,Debras2019}. Overall, the total heavy-element mass
in Jupiter can range between $\sim$ 20--60 M$_{\oplus}$ \citep{Helled2019}. 

\begin{figure}
\centerline{\includegraphics[width=1.03\textwidth]{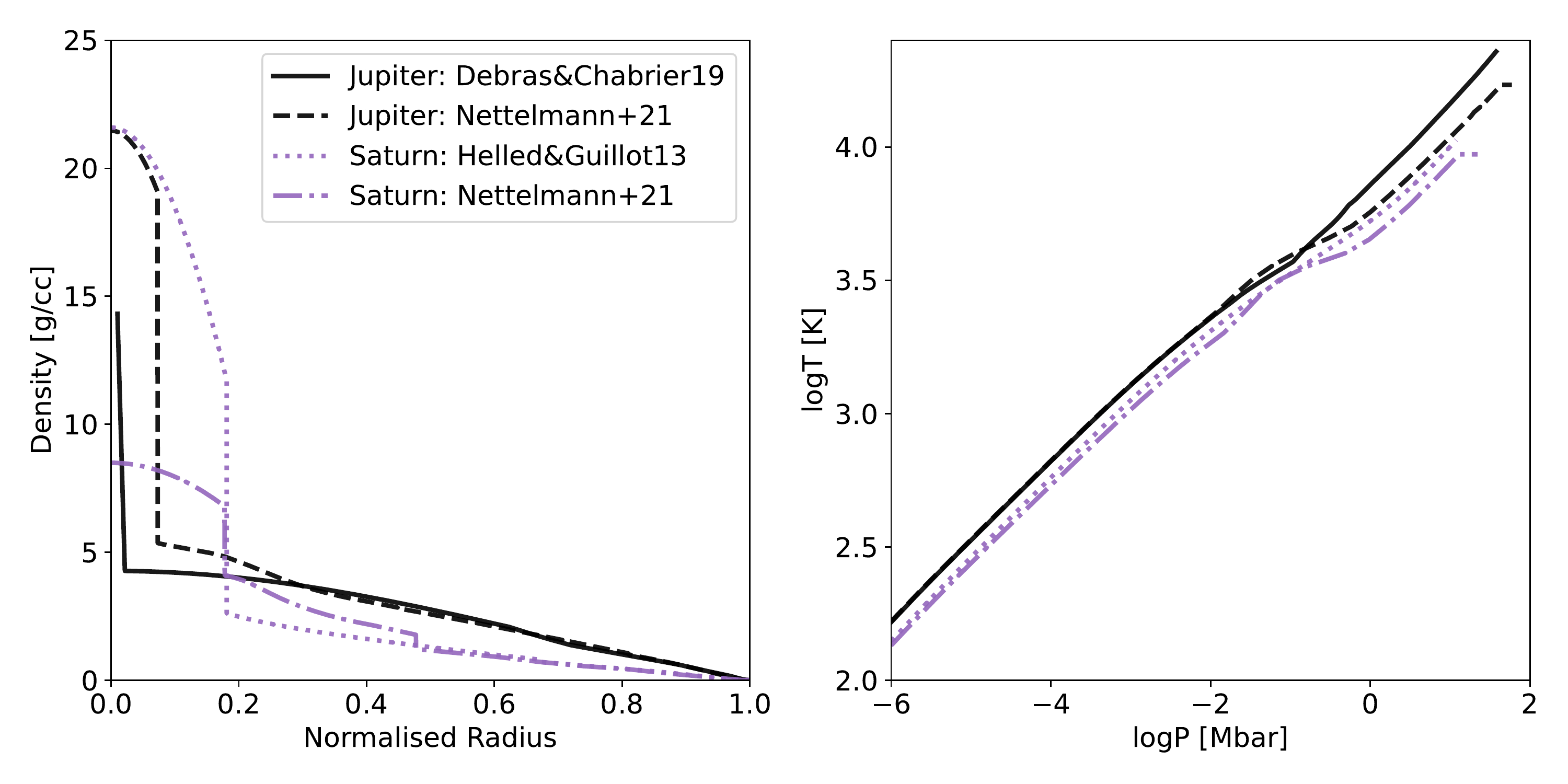}}
\caption{Examples of Jupiter (black) and Saturn (purple) structure models. \emph{Left}: Density as a
function of normalized radius. The different curves correspond to various models
as indicated in the top-right panel of the figure. \emph{Right}:  Log-temperature
vs.~log-pressure of the models.}
\label{fig:jupiter_models}
\end{figure}

Interestingly, it was recently shown that the high-order $J_n$ of the Jupiter
models fall along the same line in $J_n$--$J_{n+2}$ space, regardless of
detailed model assumptions and the H-He EOS used
\citep{Guillot2018,Nettelmann2021}. Nettelmann et al.~\cite{Nettelmann2021}
found that $J_6$ stands out in that it is neither adjusted, as $J_2$ and $J_4$
are, nor insensitive to interior model assumptions, as the higher order $J_n$
$n\geq 8$ are. As the wind contribution on Jupiter is small, $J_6$ further
constrain the interior. For instance, it can be used to put limits on the
transition pressure between the outer, He-poor outer  envelope and the inner,
He-rich envelope, where a best match was found for transition pressures of
2--2.5 Mbars \citep{Nettelmann2021}, which lies within the 0.9 Mbar--6 Mbar
demixing region in Jupiter as inferred from reflectivity measurements
\citep{Brygoo2021}. It was also shown that heavy-element gradients in the deep interior
below $\sim$ 20 Mbar can lead to high metallicities of up to $Z=0.5$ at the
compact core-mantle boundary, yielding a largely homogeneous-in-Z envelope atop.
However, the atmospheric heavy-element abundance in recent adiabatic models with
CMS-19 EoS was found to be substantially less than 1x solar, the $1\sigma$ lower
limit of the equatorial water abundance observed by {\it Juno} \citep{Li2020a},
suggesting that an adiabatic profile overestimates the density especially in the
molecular envelope where $J_4$ is sensitive, see Figure \ref{fig:nn_contribJ2n}. 
Representative density profiles and pressure-temperatures of Jupiter are shown in Fig.~4 in black. The solid and dashed curves correspond to Jupiter models as inferred by \citep{Debras2019} and  
\citep{Nettelmann2021}. 


\section{Saturn\label{sec:saturn}}
Modeling Saturn's interior is also challenging. Saturn is brighter than
predicted by fully convective models, suggesting that the effect of He rain
and/or non-convective regions is more profound in Saturn. In addition, Saturn's
rotation rate is not as well determined. Several lines of evidence sees to be
converging on an answer of about of 10 hours 33 minutes (see review by
\cite{Fortney2018} for details).
Below, we summarize a few interior models of Saturn that fit Cassini gravity
data. A detail review on Saturn can be found in \citep{Fortney2018a} and
references therein.

Helled \& Guillot \citep{Helled2013} presented 3-layer models of Saturn with a
large range of model parameters
using the SCVH EoS. 
For the range of different model assumptions, the inferred core mass and
heavy-element mass in the envelope were found to range between $\sim $5 -- 20
M$_{\oplus}$ and 0--7 M$_{\oplus}$, respectively.


Leconte \& Chabrier \citep{Leconte2013} investigated the possibility of
double-diffusive convection in Saturn and suggested that
Saturn's interior could be dominated by composition gradients (and be
non-adiabatic). 
Saturn's core mass was inferred to be between 10 and 21 M$_{\oplus}$ while the
heavy-element mass is the envelope was found to have the range  10 -- 36 $\mE$. 
This solution for Saturn does not only fit the gravity data, but can also
explain Saturn's low luminosity due to the less efficient cooling of the
planet. Since in this scenario heat transport  (cooling) is less efficient, the
deep interior can be much hotter, and the planet can accommodate larger
amounts of heavy elements.

Militzer et al.~\citep{Militzer2019} presented recent structure models of Saturn
assuming distinct layers using Monte Carlo sampling, and considering different
rotation periods and core sizes. It was found that Saturn's core mass varies
between 15 and 18~$\mE$ and that the heavy-element mass in the envelope is
small, less than $5\unit{\mE}$. 
A similar result, when using a 4-layer model of Saturn was presented by Ni
\citet{Ni2020}, where the total heavy-element mass in Saturn was found to be 12
-- 18~$\mE$. Here Saturn's interior was assumed to consist of a
compact core surrounded by a fuzzy core.

Recent Saturn models have been also been presented by Mankovich \& Fuller
\cite{Mankovich2021}. Here the interior structure models were designed to fit
not just the gravitational moments ($J_2$, $J_4$ and $J_6$) but also the the
$m=-2$ oscillation mode deduced from ring seismology. These structure models
imply that Saturn has a large region that is stable against convection extending
to $\about{60\%}$ of Saturn's radius, the interior consists of composition
gradients and that its central density is moderately low,
$\about{6}\unit{g/cm^3}$ implying that also
Saturn's core is ``fuzzy/dilute''.




Nettelmann et al.~\citep{Nettelmann2021} explored Saturn's interior and compared
the results when using different H-He EoSs. It was found that Saturn models with
the CMS EoS result in enriched envelope that extends to $< 0.4\: R_{\rm Sat}$
and compact heavy-element cores.  Saturn was found to have a fuzzy core when
accounting for EoS perturbations: then the  core extends to $\sim 0.4 R_{\rm
Sat}$, a result that is more consistent with the solution of
\cite{Mankovich2021}.
As discussed in detail in \citep{Nettelmann2021}, such a structure model
corresponds to a dilute core and helium rain, which together result in a large
region in Saturn's deep interior that is inhomogeneous in composition and is
stable against convection.

Examples of Saturn's density profiles and pressure-temperature profiles are
shown in Fig.~4 in purple. The dotted and dashed-dotted curves correspond to Saturn models presented by \citep{Helled2013} 
\citep{Nettelmann2021}. 


\section{Composition-agnostic Models}
\begin{figure}[tb!]
\centerline{
\includegraphics[width=0.5\textwidth]{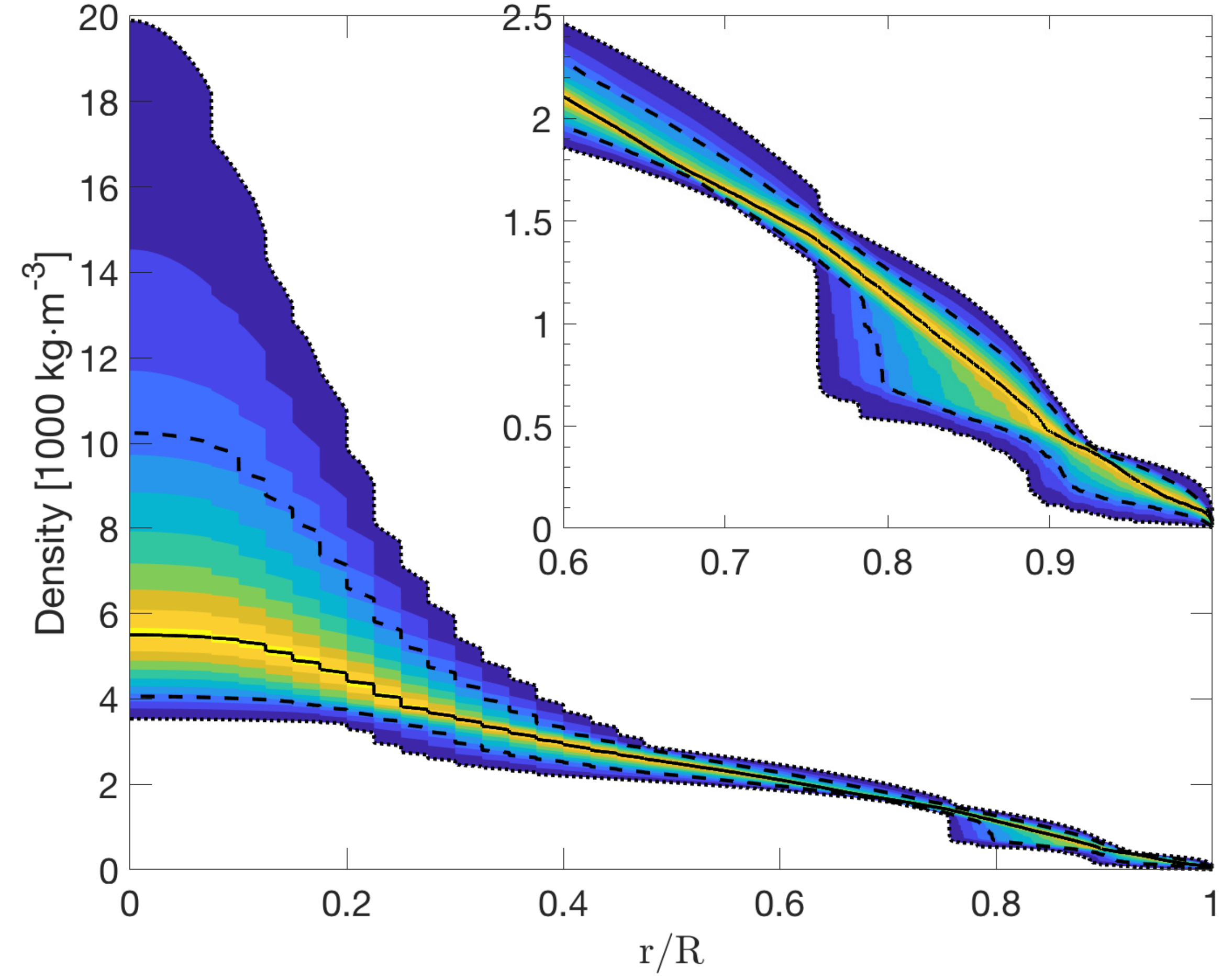}
\hspace*{4pt}
\includegraphics[width=0.5\textwidth]
{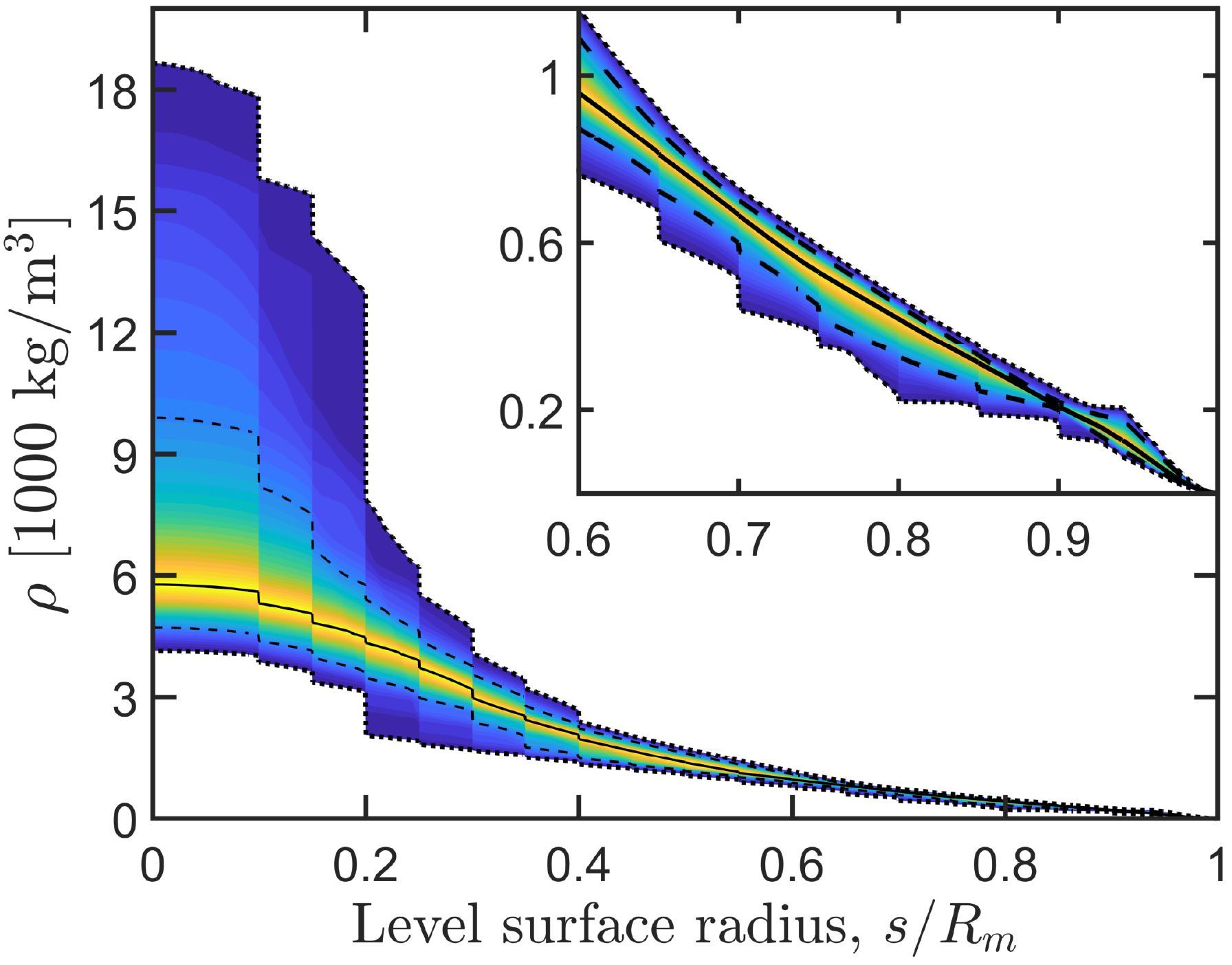}
}
\caption{{\bf Left:} Distribution of density profiles for Jupiter based on
$8^{\text{th}}$-degree polynomials from \citep{Neuenschwander2021}. The black
line marks the sample-median and the dashed lines the width of the one-sigma
deviation. The color visualizes the sample distribution and comprises $\sim96\%$
of all solutions. The polynomial-based profiles allow for up to two density
jumps and have the same precision as the polytropic-based density structures.
{\bf Right:} The posterior probability distribution of Saturn's density profiles
inferred by \citep{Movshovitz2020}. The thick black line is the sample-median of
density on each level surface. The dashed lines mark the the 16th and 84th
percentiles and the dotted lines mark the 2nd and 98th percentiles; between the
lines percentile value is indicated by color.\label{fig:emp}}
\end{figure}

A different approach for interior modeling takes a more unbiased view on the
planetary internal structure. This is by producing the so-called empirical
structure models. In this case the planetary density profile is represented by a
mathematical function, or via a series of random steps in density. Appropriate
function for giant planet interiors include polynomials and polytropes. Then, by
using the empirical representation for the density profile, all the profiles
that match the observational constraint are inferred, and a solution for the
pressure-density relation is found. From such models, the planetary composition
can be indirectly inferred by searching for mixtures that can reproduce the
density-pressure profile for an assumed temperature gradient using physical EoS.

The strength of empirical structure models is that they do not require knowledge
of the behavior of elements at high pressures and temperatures, i.e., the EoS of
the assumed materials. Also, they can probe solutions that are missed by the
standard models, in particular, solutions that represent more complex interiors
with various temperature profiles including sub- and super-adiabatic.

Recent empirical models of Jupiter were presented by Neuenschwander et
al.~\citep{Neuenschwander2021}. This study focused on the  relation between the
normalized moment of inertia and the gravitational moments using empirical
density profiles represented by (up to) three polytropes. It was shown  that
models with a density discontinuity at $\sim$ 1 Mbar, as predicted by H-He phase
separation simulations, correspond to a fuzzy core in Jupiter.

For Saturn, Movshovitz et al.~\citep{Movshovitz2020} applied a Bayesian
MCMC-driven approach to explore the full range of possible density distributions
in Saturn with density profiles parameterized as piecewise polynomial. These
models exhibit densities in the outer parts of the planet suggesting significant
heavy element enrichment (as expected), and while the inner half of Saturn was
less well constrained most models exhibit significant density enhancement
(interpreted as a core) but with density values consistent with dilute, rather
than compact heavy-element core. Some density profiles as inferred by these
studies are presented in Fig.~5.


\newpage
\section{Winds on Jupiter and Saturn}\label{sec:diffrot}

Jupiter's visible atmosphere is structured into darker belts and brighter zones,  which can be related to the winds that are blowing around the planet both in prograde and in retrograde direction with respect to the planet's steady rotation. Strong zonal winds are also observed on Saturn. There are also upwellings  and downdrafts through which heat from the deep interior is transported outward. Constraining the depth of the winds on Jupiter and Saturn was one of the main drivers for the Juno and the Cassini Grand Finale missions, respectively. 
For a recent review on thermal,  compositional, and wind properties in the upper wind region we refer the reader to \citet{Kaspi2020,Guillot2022}.  

A depth of 1000 km corresponds to 1.3\% of the radius of Jupiter, or 1.7\% on Saturn. As can be estimated from Figure \ref{fig:nn_contribJ2n}, density perturbations that reach as few percent in radius would reflect especially on the higher-order gravitational harmonics. 
Therefore, by measuring the high-order  harmonics one obtains signatures on the depth of the winds.  In addition, any signature in the measured odd gravity harmonics is a direct signature of the winds \cite{Kaspi2013a}. Also it should be  noted that while the low-order even harmonics are dominated by the deep interior, the wind also contribute to these harmonics, and are even more significant for  $J_6$, $J_8$, and $J_{10}$ \cite{Kaspi2020}.

The observed winds on Jupiter and Saturn can be mapped from the cloud level onto cylinders that rotate parallel to the planet's spin axis \citet{Hubbard1999, Kaspi2010, Kulowski2021}. Such a vertical projection of the winds allowed Galanti et al.~(2021) \citet{Galanti2021} to obtain a much better match to the high-order even $J_n$ than for a radial projection. This view is supported by 3D hydrody\-namic simulations of zonal flows in rotating spherical shells \citep{Dietrich2021}.

By now there is general agreement that scale height of the wind decay depth is significantly less than the planet's radius,  but is much larger than the depth of water condensation.  Regardless of the assumed decay depth profile. The exact wind braking mechanism is still not perfectly  understood but is expected that the winds decay due to ohmic dissipation linked to the metallization of hydrogen \cite{Liu2008,Cao2017a}. 


While detailed models are still being developed, thanks to {\it Juno} and {\it Cassini} data the depth of the winds on the gas giants have been constrained \cite{Kaspi2020}. 
The observed gravity and magnetic field data suggest that the depths of the flows on Jupiter and Saturn are $\sim 3000$ km and $\sim 9000$ km, respectively \citep{Galanti2020, Kaspi2020}. 
 Nevertheless, further work is needed  to better resolve the detailed wind structure and braking mechanism. It  is also desirable to establish comprehensive giant planet models where the three aspects of   flows, magnetic field, and internal structure are considered.  


\newpage
\section{Love numbers}\label{sec:knm}

The tesseral harmonics $C_{nm}$ probe axial-asymmetric perturbations of the planetary gravity field. Such can arise from tides \citep{Wahl2017b} or other effects.
Due to their periodic nature, the tidal influence on the $C_{nm}$ can rather easily by extracted from the observed signal. The residuals will then be a probe of normal modes or other effects (see Iess et al (2018) \citet{Iess2018} for Jupiter and Iess et al (2019) \citet{Iess2019} for Saturn).

Planetary tidal response is commonly parameterized by the Love numbers $k_{nm}$, which are the linear response coefficients between the induced planetary gravitational field perturbation at the planet's sub-satellite point $a_s$, and the gravity field of the perturber there, $W(a_s)$. If the satellite orbits in the equatorial plane, $a_s=a_0$. 
Writing the planetary gravitational $V$ as:
\begin{equation}\label{eq:V0rtt}
	V(r,\,\varphi,\,\theta) = V_0(r) + \sum_{n=2}^{\infty}V_n + \sum_{n=2}^{\infty}V_n
	+ \sum_{n=1}\sum_{m=1}^n V_{nm},
\end{equation}
where $V_0$ is the spherically symmetric part, and the gravity field $W$ of the tide-raisig perturber as 
\begin{equation}\label{eq:WWnWnm}
	W = \sum_{n=0}^{\infty} W_n + \sum _{n=1}^{\infty}\sum_{m=1}^n W_{nm}
\end{equation}
one can define the Love numbers of a fluid body as $k_{nm}=V_{nm}/W_{nm}$.
If the planet resides in a (equilibrium)  state of 1:1 spin-orbit resonance with its tidal perturber so that the tidal buldge raised by the satellite on the planet is static in the frame corotating with the planet, $_{knm}$ can be obtained from expression (\ref{eq:V3dim_Jn0}) and written as:  
\begin{equation}
	k_{nm}^{static} = - \frac{3}{2}\:\frac{(n+m)!}{(n-m)!}\:
	\frac{C_{nm}}{q_{tid}\:P_n^m(\mu_S)}\:\left(\frac{r_S}{R_{eq}}\right)^{n-2}\:.
\end{equation}
and be computed from an interior model \cite{Wahl2017b}. 

In general, tides are a dynamic phenomenon. The tidal buldge floats around the planet, and to Coriolis force acts on that flow \citep{Idini2021}.
Similar to the $J_{n}$--wind effect, one may write: 
\begin{equation}
	k_{nm} = k_{nm}^{static} + \Delta k_{nm}^{dyn}. 
\end{equation}
Cassini \cite{Lainey2017} and Juno  measurements \citep{Durante2020} revealed that the observed $k_{22}$ values are close to their static values, which for Jupiter is  $k_2=0.5897$ \cite{Nettelmann2019, Wahl2020} with an uncertainty of less than 0.02\%, as the underlying interior models are already well constrained by the precisely measured $J_2$ and $J_4$ values. 
By analyzing 10 Juno passes, \cite{Durante2020} were able to reduce the $3\sigma$ uncertainty in $k_{22}$ to only 3\%, revealing a clear deviation of 1--7\% from the static value.
Dynamic contributions to the Love numbers have thus observationally been found to exist on Jupiter \citep{Idini2021}, however, their quantitative computation is challenging can can arise from different physical origins.


Consequently, Idini and Stevenson \cite{Idini2021} as well as Lai \cite{Lai2021} aimed to explain the observed dynamic contributions. Idini and Stevenson \cite{Idini2021} showed that including the tidal flow (but ignoring the Coriolis force) increases $k_{22}$ by $\sim$13\% with respect to the static value of a non-rotating $n=1$ polytrope. Including Coriolis adds a negative dynamic correction bringing the observed value (0.565) into agreement with the superposition of $k_2^{static}+k_2^{dyn}$. Interestingly, the same procedure applied to $k_{42}$ yields a correction of an  opposite sign to what has been observed \cite{Idini2021,Lai2021}). 

To some extend, the $k_{nm}$ 'measurements' rely on simplifications made in the fitting procedure of Juno's trajectory. The reported values \cite{Durante2020} are based on satellite-independent $k_{nm}$ values, while it has been shown \cite{Wahl2017b,Nettelmann2019,Wahl2020}
that the static $k_{nm}$ differ as a function of the tidal forcing parameter $q_{\rm tid}=-3(M_S/M_p)/(Req/a_S)$, which differs among satellites of different masses $M_S$ and orbital distances $a_S$. The observational determination of Jupiter's tidal response coefficients is still ongoing and is a key goal of Juno's extended mission. Further moments showing signs of dynamic influence are $k_{42}$ and $k_{44}$ \cite{Idini2021}. 

Love number observations provide a new opportunity to further constrain the internal structure of giant planets. As discussed above, both Jupiter and Saturn could posses thick stably stratified deep interiors and further stable regions in the envelope where g-modes can exist. If they fall in resonance with the tidal forcing frequency they can significantly influence the tidal response. Since broadening and shifts of the resonance depend on the location and width of the stably stratified region \cite{Lai2021}, the tidal response enhancement can even occur out-of-resonance \cite{Lai2021}. This effect enables us  to investigate the width and location of possible stratified layers.


\section{Summary \& Outlook}
Giant planets are key objects to characterize as their composition and internal
structure reveal key information for formation theories and on the conditions of
the protoplanetary disks where the planets formed.

\begin{figure}
\centerline{\includegraphics[width=0.85\textwidth]{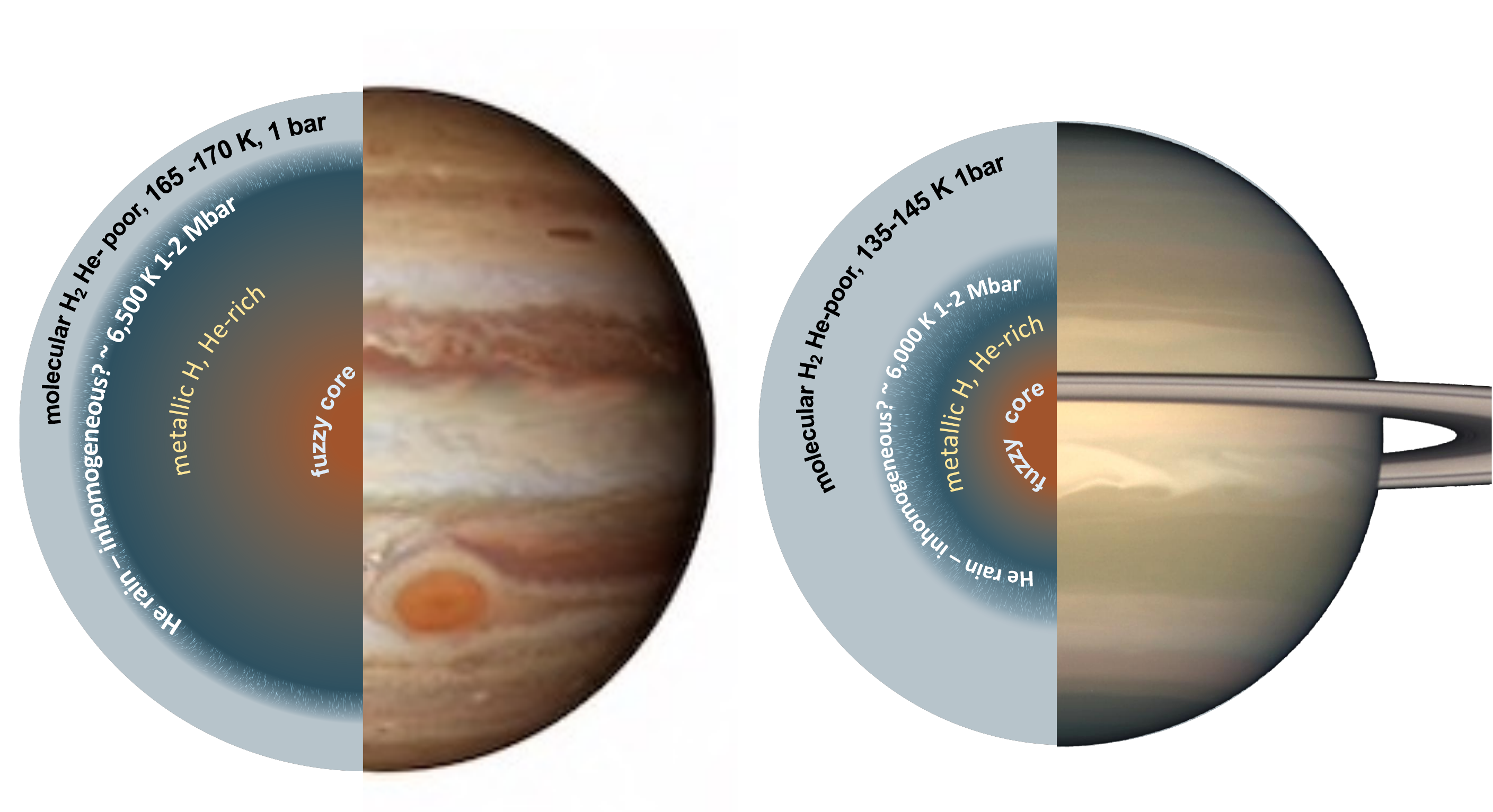}}
\caption{Sketches of the internal structures of Jupiter and Saturn. The outer layers of both planets is expected to mostly 
consist of molecular hydrogen. Due to the process of helium rain the outer envelope is depleted in
helium, and there is a region where helium rain takes place. The inner part of
the planets consists of metallic hydrogen with enriched helium, and possibly
composition gradients and/or fuzzy core that can extend to more than half of the
planetary radius (see text for further details). The exact heavy-element mass and its distribution is still being investigated. Figure modified from \cite{Helled2019}. }
\label{fig:interior_sketches}
\end{figure}

Our understanding of giant planet interiors has been revolutionized in the last
several years thanks to the accurate data from the {\it Juno} and {\it Cassini}
missions. It is now clear that both planets have complex interiors that include
non-convective regions, composition gradients and fuzzy cores.
Figure~\ref{fig:interior_sketches} provides a schematic representation of the
internal structures of Jupiter and Saturn. While great progress has been made, new questions have arisen and much work and future investigations are
required. 

Such future investigations may include: 
\begin{itemize}
\item A detailed comparative planetology should be obtained by
having similar observables for Jupiter and Saturn, such as has been achieved for
the gravity and wind velocity measurements. For example, at present, only for Jupiter are atmospheric elemental abundances  
measured below the radiative-convective boundary.  
Their vertical abundance profiles are particularly valuable as these
elements do not condense. 
Therefore a probe to directly measure the atmospheric
composition of Saturn is clearly desirable. Such a probe can determine the He 
abundance in Saturn's atmosphere, constraining the magnitude of He rain, and determining Saturn's atmospheric metallicity. 
\item Jovian seismology could further constrain the planetary structure. For example, the detection and characterization of 
 planetary f/g-modes as been  
detected in Saturn's rings, doppler imaging, and measurements of the dynamic
parts of the Love numbers. In that regard, Juno extended mission
is on its way to let us look deep into Jupiter's interior of Jupiter via improved  gravity
measurements.
\item The calculated structure depends on the used EoS as small differences in
the EoS can lead to relatively large differences in the inferred composition. As
a result, it is clear that further investigations (both theoretically and
experimentally) of materials at planetary conditions are desirable.  
\item A detailed comparison between Jupiter and Saturn and giant exoplanets. Using the observed trends and inferred compositions and structure of a large number of giant exoplanets can be used to put Jupiter and Saturn in perspective, and at the same time the detailed investigation of the solar system giant planets is key to test and assess the nature of gas giant planets.    
\end{itemize}

Overall, it seems that current models of Jupiter and Saturn have cannot reproduce all the observed measurements at once.  Interior models of both Jupiter and Saturn  typically predict relatively 
low envelope metallicities: solar or sub-solar,  which is at odds with
observations \cite{Atreya2016,Guillot2022}.  This could imply that the measured atmospheric metallicities do
not represent the bulk composition of the outer envelope and/or that our models are based on inappropriate assumptions/steups such as the applied
EoS, the entropy assigned, or the assumed temperature profile. We hope that future
investigations will reconcile this issue sand will lead to a more comprehensive understanding of Jupiter and Saturn.

\end{document}